# Growth Rates of Knowledge

Nick Zhang
nick@iwuzhen.org

**Abstract**

This is an evolving document. It is devoted to summarizing patterns and laws of knowledge growth. By examining a variety of parameters in data sources such as Wikipedia and Microsoft Academic Graph, we can get deeper insights of how knowledge evolves.

**Introduction**

Wikipedia has reached the level of Britannica in quality in 2005 [Giles,2005]. It can be used as a proper proxy of knowledge. Studying the evolution of Wikipedia can help us understand growth of knowledge. The specialization of knowledge is caused by knowledge burden and limitations of human cognitive faculty [Jones,2009]. There is no renaissance man any more. We increasingly rely on computers to maintain and process knowledge.

We have collected monthly snapshots of Wikipedia since its inception of 2001, all the way to January of 2022, and made a quantitative analysis in order to model its growth. We only focus on the English edition of Wikipedia as it has the highest quality and quantity.

Wikipedia is most often used with keywords search. In 2004, a hierarchical architecture called category is introduced in order to organize contents in knowledge ontology. For example, 'Linear algebra' is a subcategory of a larger category 'Algebra', which in turn is a subcategory of 'Mathematics', then a subcategory of 'Formal sciences', and etc. This classification scheme has long been utilized in knowledge representation and typing in programming languages. An article or a category can have multiple parent categories, for instance, 'Biochemistry' is a subcategory of both 'Biology' and 'Chemistry', similar to multiple-inheritance of types in Object-Oriented programming. The strict ontology hierarchy is an algebraic structure called *lattice* instead of tree which does not have multiple inheritance [Sowa,1984]. Wikipedia's editorial process is too democratic and lacks of rigor, therefore, loops can happen, which is forbidden in lattice. For example, in Wikipedia, 'Physics' and 'Mathematics' are children of each other. At present, circulations of this kind have not caused great harms yet. An informal but convenient hierarchical structure can serve knowledge graph with simple and usable ontology. We analyze not only articles, but also categories and their structures.

By 2020, Microsoft Academic Graph (MAG, formerly MAS) has collected more than 250 million academic publications, mainly papers with some books and patents [Sinha, 2015]. Concept detection and taxonomy learning techniques are used to establish ontology hierarchy automatically [Wang, 2019]. The entire MAG is classified with FOS ("Fields of Study"). At the top there are 19 fields, such as, 'Mathematics', 'Physics', 'Engineering', 'Political

Science' and so on. An MAG field is similar to Wikipedia category, and a field may contain multiple subfields.

In order to compare Wikipedia with MAG fairly, we have further selected a dozen or so academic categories from Wikipedia, such as 'Mathematics', 'Physics', 'Chemistry', 'Geology', 'Computer science', 'Engineering', 'Psychology', 'Geography', 'Sociology', 'Political sciences', 'Philosophy', and etc., to form what we call Wikipedia Academic Group (WAG). In general, WAG is of higher quality than the average Wikipedia.

**Growth Rates of Knowledge**

If the total volume of knowledge is delineated by the number of papers, it is commonly agreed that it grows exponentially [Fortunato,2018, Wang,2021]. The dramatic growth of publications does not necessarily reflect the real expansion of knowledge boundaries. Knowledge can decay too, sometimes exponentially [Arbesman,2013]. If lexical diversity (new unique phrases in paper titles) is used instead as a measure of cognitive extent, then the growth tends to be linear. Hence, it was also argued that the space of ideas (cognitive content) grows linearly, or logarithmically with the number of academic publications [Milojević,2015].

The numbers of articles and categories of Wikipedia, categories of WAG (Wikipedia Academic Group) and fields of MAG can be used to quantify the growth of knowledge as well. They grow at $O(\frac{t}{\ln(t)})$, $O(t*\ln(t))$, $O(t))$ and $O(t*\ln(t))$, respectively. Ideas (categories and fields) are cheap as they grow faster than the real Wikipedia articles. MAG papers are churned out even cheaper as the average quality of papers drops significantly if quality could be measured by the disruption index [Park,2021, Wang,2021].

We have distinguished Early Wikipedia (before 2006) from Wikipedia as they follow different growth trajectories. Early Wikipedia grew dramatically faster, as in that period, it could be relatively easy to jump start initial data from other extant data sources. Wikipedia has become more mature in 2006. Early Wikipedia can be characterized as exogenous growth, whereas, Wikipedia after 2006 as endogenous growth.

The monthly increment of the endogenous period of Wikipedia after 2006 can be well fitted by the reciprocal of logarithm:

$$mi = \frac{140000}{ln(t)}$$

The logarithmic integral, $li(t) = \int \frac{1}{ln(t)} dt$ can be approximated by $\frac{t}{\ln(t)}\left(\left(1 + \frac{1}{\ln(t)} + \frac{3}{(\ln(t))^2}\right)\right)$ or even more simply, $\frac{t}{\ln(t)}$.

Then the entire growth curve can be fitted with
$$A = 140000*li(t) + 1,350,000$$

where 1,350,000 is just the total number of articles in 2006. It increments by almost 140,000/ln (t) every month. Applying the fitting formular to the endogenous period of Wikipedia, the monthly error is only about 4% in 2007, and within 0.7% after 2010. We can confidently predict the growth of Wikipedia with high accuracy.

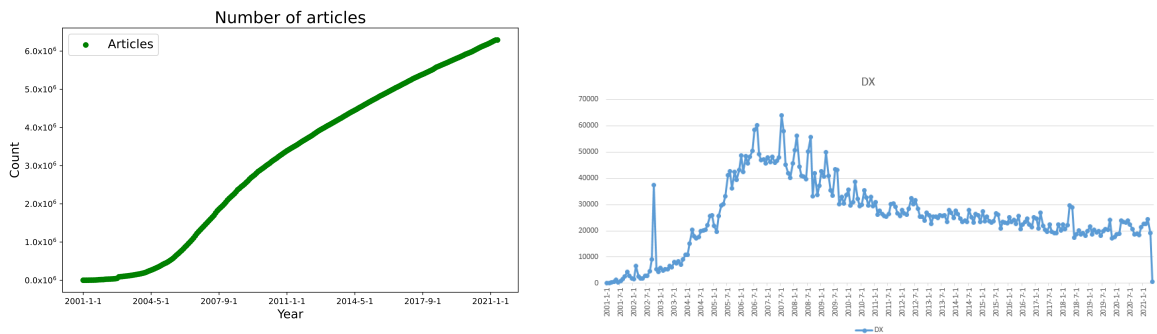

Figure 1. The growth of Wikipedia since inception

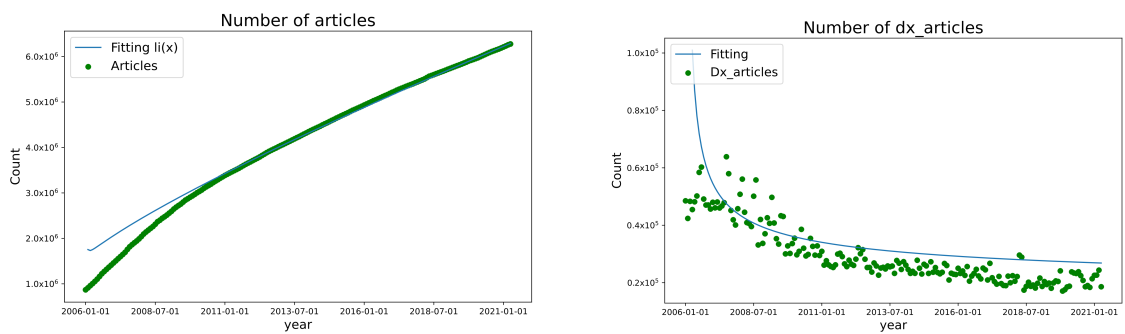

Figure 2. The growth of endogenous period of Wikipedia after 2006

It is generally believed Moore's law and Koomey's law are restricted by people's cleverness. Moore's statement is interesting: "No exponential is forever: but 'forever' can be delayed!". Nothing can sustain the exponential growth, even human knowledge. This dampening effect can be better characterized by Nick's law, which also happens to be logarithmic integral [Zhang, 2022].

It is worth noting that the number of articles, the length and the number of editors of Wikipedia can all be approximated by logarithmic integral. The number of Wikipedia categories and the number of MAG fields follow $t*\ln(t)$. It is observed that the growth curve of lexical diversity might be more tightly fitted with logarithmic integral instead of linearity as the tail shows gradual slowdown.

In exogenous period, knowledge grows faster, like easier to collect low-hanging fruits, while in endogenous period, it is getting harder and harder. There were attempts to show that scientific discoveries obey a simple law of exponential decay [Arbesman,2011] but the assumption that the total amount of knowledge has an upper limit seems arbitrary.

**The Table of Growth Rates**

The following table summarizes the changes over time of different parameters in Wikipedia, Early Wikipedia, Wikipedia Academic Group and MAG. BA Network is listed as well for comparison. The methodologies are supplemented in the appendices.

| Parameters | Wikipedia | Early Wikipedia (2006 and before) | Wikipedia Academic Group | BA Network | MAG |
|---|---|---|---|---|---|
| nodes, articles, papers, N | t/ln(t) | cubic | t/lnt | t | sub-exponential $O(2^{t/lnt})$ |
| edges, links, citations, references, E | t | ~cubic | t | t | sub-exponential $O(2^{t/lnt})$ |
| category, fields | tln(t) | ~cubic | t | | tln(t) |
| degree (total degree/N) directed graph: E/N | quasi-linear | quasi-linear | quasi-linear, can fit with lnt | M | quasi-linear |
| length, size, number of words | t/ln(t) | polynomial | t/ln(t) | | |
| average length, length/N | sublinear | quasi-linear | sublinear, tend to be flat in the end | - | - |
| number of edits | t | polynomial | t | - | - |
| average edits: edits/N | lnt | - | lnt | - | - |
| number of editors | t/ln(t) ~ t | polynomial | t/lnt ~ t | - | - |
| average editors | ln(t) | - | ln(t) | - | - |
| average edits by editors (edits/editors) | constant | - | constant | - | - |
| degree distribution entropy | $(A/c^2)*x[ln(N)^2-2ln(N)+2]$+constant | | $(A/c^2)*x[ln(N)^2-2ln(N)+2]$+constant | constant, | |

| normal structural entropy | links-in seems slightly upward. links-out tends to be constant. hard fields tend to have higher normal structural entropy | | links-in seems slightly upward. links-out tends to be constant | fits with ln(N)+c | citation links-in is downward. references links-out is upward |
|---|---|---|---|---|---|
| density directed graph: E/N(N-1) | (ln(t))²/t | | (ln(t))²/t | m/t | 1/t |
| valid diameter | constant, mostly 4 some 5. | | | ln(N)/lnln(N) | |
| average shortest distance | | | | ln(N)/lnln(N) | |
| clustering coefficient | | | | ln(N)²/N | |
| power law | $P(k) \sim m^2(ln(t))^2 c^{-2} k^{-3}$ | | | $P(k) \sim 2m^2 k^{-3}$ | |

**Conclusions**

In Appendix 6, we demonstrate that Wikipedia is more effective in quantifying knowledge growth. Both Wikipedia itself and academic papers collected by Wikipedia seems to obey similar patterns. They are prescribed with two phases of growth. Early exogenous phase increases very fast, whereas normal endogenous phase follows quasi-linear trajectory and can often be fitted with logarithmic integral.

# References

Arbesman, S. Quantifying the ease of scientific discovery. *Scientometrics* **86,** 245–250 (2011).

Arbesman, S. *The Half-Life of Facts: Why Everything We Know Has an Expiration Date* (2013)

Fortunato, S., et al., Science of science. *Science*, **359** March, (2018).

Funk, R.J. & Owen-Smith J. A dynamic network measure of technological change. *Management Science*, 63, 791–817 (2017).

Giles, J. Internet encyclopaedias go head to head, *Nature* **438,** 900–901 (2005).

Jones, B. F. The burden of knowledge and the "death of the renaissance man: Is innovation getting harder. *Rev. Econ. Stud*. 76, 283–317 (2009).

Milojević, S. Quantifying the cognitive extent of science. *J. Informetrics*. **9**, 962–973 (2015).

Park, M., Leahey, E. and Funk, R. Dynamics of Disruption in Science and Technology, arxiv:2106.11184v2 (2021)

Sinha, A., Shen Z., Song Y., Ma H., Eide D., Hsu B., & Wang K. An Overview of Microsoft Academic Service (MAS) and Applications. *Proceedings of the 24th International Conference on World Wide Web (WWW '15 Companion)*. ACM, New York, NY, USA, 243-246. DOI=http://dx.doi.org/10.1145/2740908.2742839 (2015).

Sowa, J. 1984. *Conceptual Structures - Information Processing in Mind and Machine*, Addison-Wesley (1984)

Wang, D. & Barabási A., *The Science of Science*. (2021)

Wang, K., et al, A Review of Microsoft Academic Services for Science of Science Studies, *Frontiers in Big Data*, vol.2 December (2019).

Wu, L., Wang D. & Evans J. Large teams develop and small teams disrupt science and technology, *Nature*, (2019)

Zhang, N. Moore's Law is dead, long live Moore's Law! arxiv: 2205.15011 (2022)

**Appendix 1. Number of Articles of Wikipedia**

It can be seen that logarithmic integral almost perfectly fits the data of the endogenous period of Wikipedia after 2006. MAPE error only reached 4% in 2007, and all errors were within 0.7% after 2010.

|  | Fitting | MAPE |
|---|---|---|
| 2021-5-1 | 6304698 | -0.0053 |
| 2021-7-1 | 6347547 | -0.0052 |
| 2021-8-1 | 6368943 | -0.0054 |
| 2021-9-1 | 6390319 | -0.0054 |
| 2021-10-1 | 6411676 | -0.0058 |
| 2021-11-1 | 6433014 | -0.0062 |
| 2021-12-1 | 6454334 | -0.0068 |
| 2022-1-1 | 6475635 | -0.0073 |
| 2022-2-1 | 6496917 | -0.0078 |

So we predict the number of articles in the next few years as below.

| 2023 January | 6,729,834 |
|---|---|
| 2024 January | 6,981,559 |
| 2025 January | 7,230,982 |
| 2026 January | 7,478,255 |

**Appendix 2. Article Size Distribution**

The growth of total size or length of Wikipedia follows logarithmic integral as well. The average size of an article grows sub-linearly. The size distribution approximates lognormal as shown in the following figure.

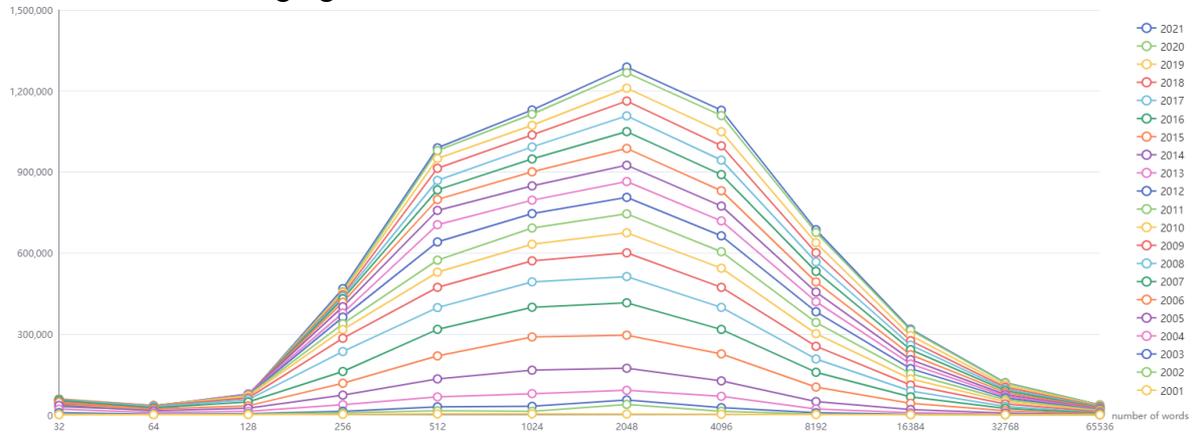

Figure 2-1. The article size follows lognormal distribution

**Appendix 3. Number of Categories of Wikipedia and Wikipedia Academic Group**

The number of categories of Wikipedia grows super-linearly as shown below (left). The monthly increment of categories since 2006 can be well fitted with $mc = 2000*ln(t)$ as shown below (right). Hence, the number of categories can be approximated with $c = 2000(t+12)ln(t+12)$.

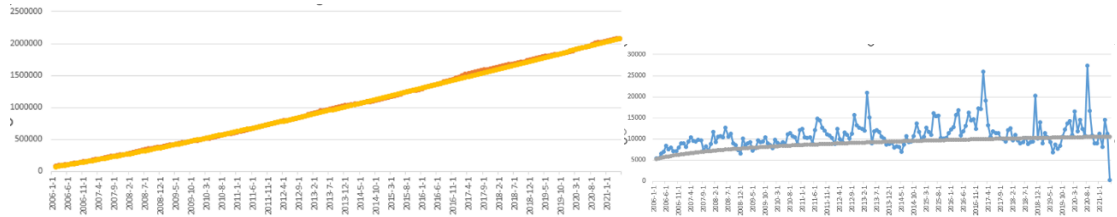

Figure 3-1 Growth of Wikipedia categories and academic group

We predict the growth of categories as following:

| 2023 January | 2,334,875 |
| --- | --- |
| 2024 January | 2,488,645 |
| 2025 January | 2,643,672 |
| 2026 January | 2,799,895 |

The academic group of Wikipedia is of higher editorial quality. We have selected a dozen or so stable academic categories, 'Mathematics', 'Physics', 'Chemistry', 'Computer science', 'Biology', 'Material science', 'Medicine', and 'Engineering', to form the academic group. All articles belonging to subcategories with three levels below the top categories are counted. The results can be fitted with linear formular:
$y = 30t + 3800, t\ is\ the\ number\ of\ months\ since\ 2007$

The editors of Wikipedia academic contents are less inclined to form new categories than average editors.

## Appendix 4. Number of Papers of MAG

The number of papers of MAG grows exponentially or quasi-exponentially as shown below (left). Its logarithmic grows in scale of logarithmic integral (right) and can be fitted with

$$y = 0.4 \frac{t}{lnt} + 19.53$$

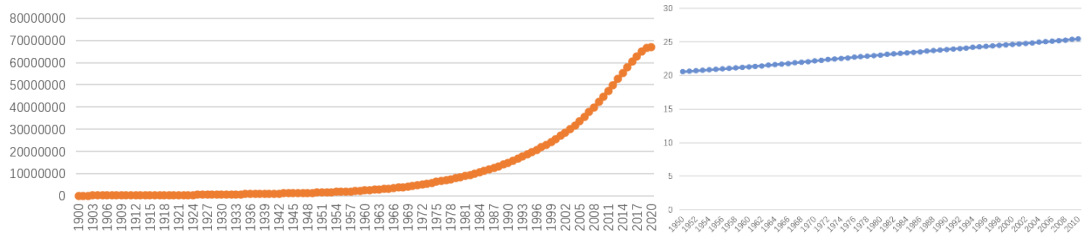

Figure 4-1 Growth of MAG papers

**Appendix 5. Number of Fields of MAG**

The number of fields of MAG grows super-linearly, similar in magnitude to the number of categories of Wikipedia. It can be fitted with $y = a*t*\ln(t) - at + b$, $a = 2467, b = 147079$. The fitting curve (gray) is shown below where blue is the real data:

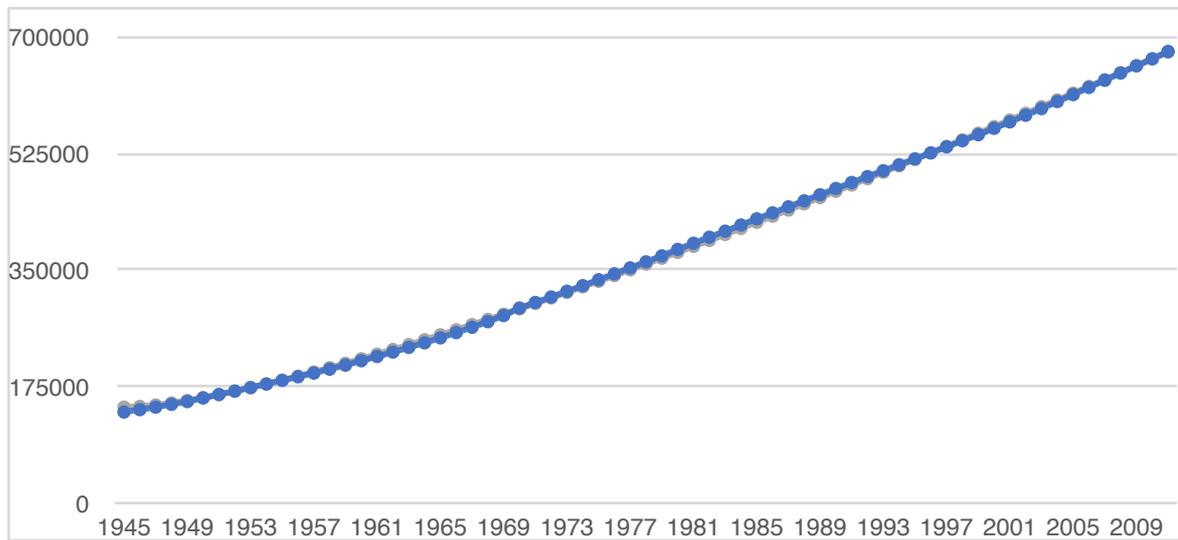

Figure 5-1. Growth Curve of MAG FOS

**Appendix 6. Disruption Index and Knowledge Growth**

The disruption index is increasingly used to measure innovativeness of scientific and technological achievements [Funk,2017, Wu,2019, Park,2021]. We argue that it may not be a universal metric. The disruption index *D* of a focal paper (blue diamond) is defined with its citation network which consists of its predecessors (gray circles) and subsequent works (rectangles), as shown in Figure 1. The naïve intuition is that subsequent works of a disruptive research are less likely to cite its predecessors. *D* ranges from -1 (least disruptive) to 1 (most disruptive).

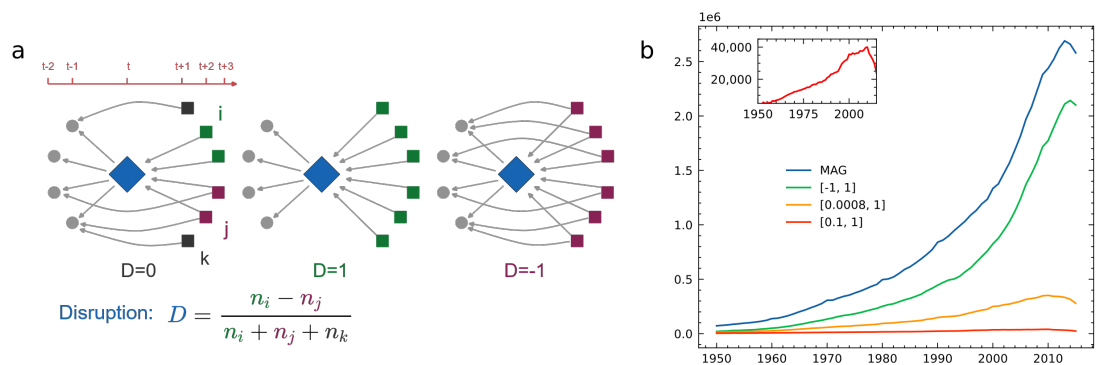

Figure 6-1. (a) The subsequent works may cite only the focal paper (green *i*); only its predecessor (black *k*); and both focal paper and predecessors (brown *j*). (b) Even as the *D* interval is reduced to [0.1,1], which results in only tens of thousands of extremely 'disruptive' papers left (inset in red), it is still cubic.

Wikipedia was founded in 2001 and began to systematically collect academic papers in 2005 in its references section. A 2022 January snapshot of English edition contains about 1.85 million papers. The meta data of these papers are cross checked with MAG, and 1.2 million papers are selected with publication dates from 1965 to 2010, and the result set is named W-1200. Another 1.2 million papers with the highest *D*s are selected from MAG and named D-1200. Papers from 2010 onwards are ignored because they have not accumulated adequate citations. All MAG papers are sorted according to citations in descending order and named C-top.

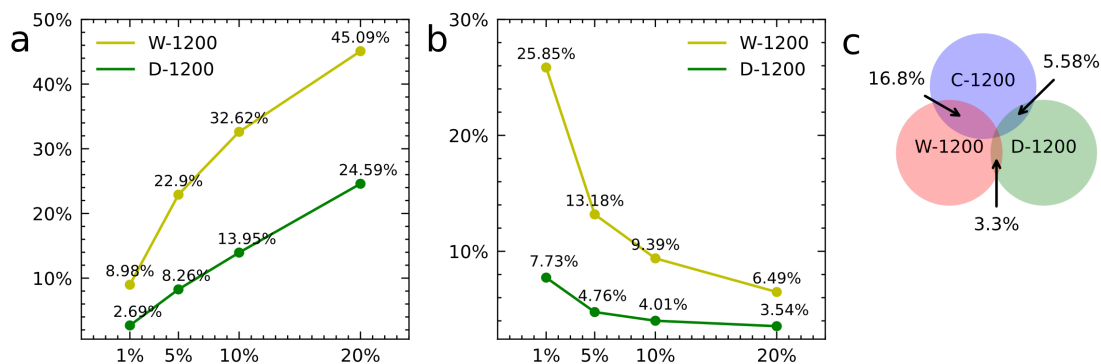

Figure 6-2. X-axis is the percentile of C-top. (a) Y-axis is percentage of the size of intersections of W-1200 and D-1200 with C-top over 1.2 million. (b) Y-axis is percentage of same intersections over the size of subsets of C-top. Top 5% of C-top is of about 2 million. (c) C-1200 (top 1.2 million from C-top) intersects with W-1200 and D-1200.

The top 5%, 10% and 20% subsets of C-top are intersected with W-1200 and D-1200, respectively. W-1200 ∩ C-top are generally 2-3 times larger than D-1200 ∩ C-top. The advantage of W-1200 over D-1200 is clear.

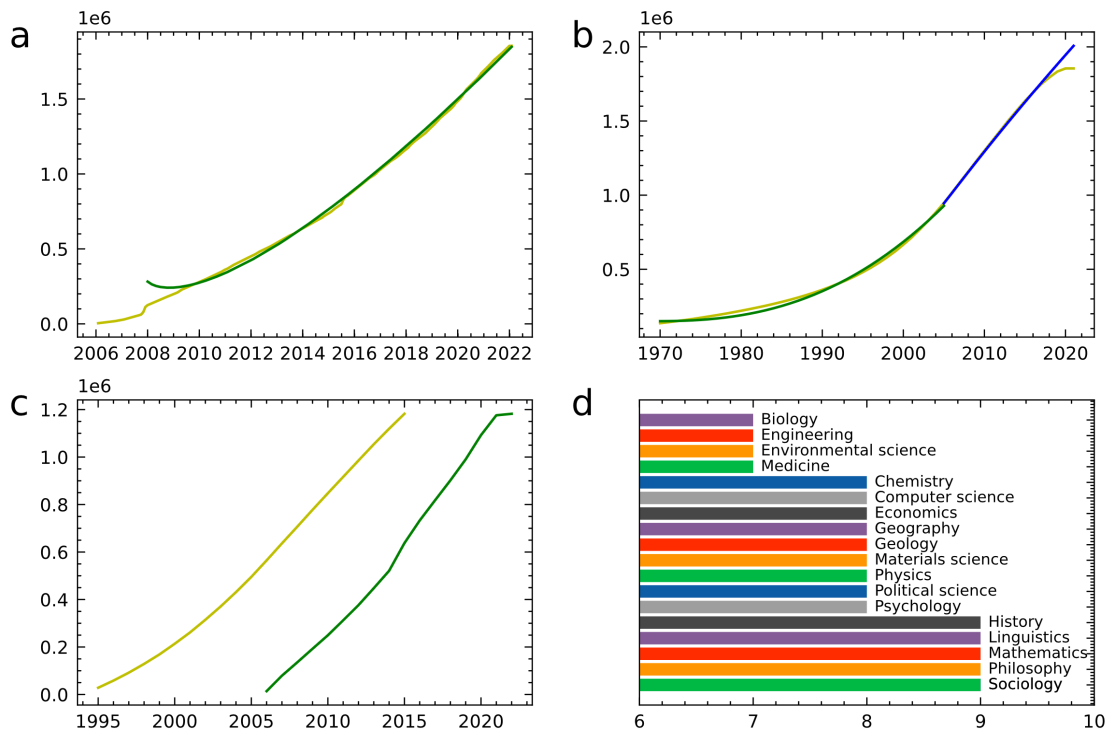

Figure 6-3. (a) The growth curve the number of papers included in Wikipedia is quasi-linear and can be fitted by: $y = 160{,}000(ct)ln(ct) + 300{,}000,$ where $c = 0.033,$ and the monthly increment by $mc = 160{,}000c\bigl(ln(ct) + 1\bigr)$. (c) The inclusion time by Wikipedia is about 8 years behind publication dates and varies with different fields (d).

Publication dates (orange) behave differently in two phases. Before 2000, the curve can be fitted with a quadratic function (green, Figure 6-3(b)). After 2000, the growth slows down and it can be more properly fitted with logarithmic integral (blue) similar to the number of articles in Wikipedia. The inclusion time by Wikipedia is about 8 years behind publication dates (Figure 6-3(c)) and varies with different fields (Figure 6-3(d)).

The most disruptive achievements would not likely be ignored by Wikipedia as its editors are more comprehensive than a small group of domain experts surveyed. It is worth noting that both inclusion time curve (Figure 6-3(a)) and publication dates curve (Figure 6-3(b)) are quasi-linear, between $O(t/ln(t))$ and $O(tln(t))$, similar in magnitude to that of Wikipedia articles, Wikipedia categories, WAG articles, WAG categories and MAG fields.